\documentclass[aip, jmp, amsmath,twocolumn,amssymb,reprint,]{revtex4}

\usepackage{gensymb,textcomp}
\usepackage{graphicx,epstopdf}
\usepackage[usenames]{color}
\usepackage{latexsym,comment}
\usepackage{hyperref}
\usepackage{bm}
\usepackage{wasysym} 

\usepackage{docs}%
\usepackage{dcolumn}

\newcommand{\comments}[1]{} 

\begin{document}
\title{Electrical activation and electron spin resonance measurements of implanted bismuth in isotopically enriched silicon-28}

\author{C. D. Weis$^1$}
\email[Please contact the corresponding author under ]{cdweis@lbl.gov}
\author{C. C. Lo$^{1,2}$}
\author{V. Lang$^3$}
\author{A. M. Tyryshkin$^4$}
\author{R. E. George$^5$}
\author{K. M. Yu$^6$}
\author{J. Bokor$^{2}$}
\author{S. A. Lyon$^4$}
\author{J. J. L. Morton$^{3,5}$}
\author{T. Schenkel$^{1}$}
\affiliation{$^1$Accelerator and Fusion Research Division, Lawrence Berkeley National Laboratory, Berkeley, California 94720, USA}
\affiliation{$^2$Department of Electrical Engineering and Computer Sciences, University of California, Berkeley, California 94720, USA}
\affiliation{$^3$Department of Materials, University of Oxford, Oxford OX1 3PH, United Kingdom}
\affiliation{$^4$Department of Electrical Engineering, Princeton University, Princeton, New Jersey 08544, USA}
\affiliation{$^5$CAESR, Clarendon Laboratory, Department of Physics, University of Oxford, Oxford OX1 3PU, United Kingdom}
\affiliation{$^6$Materials Sciences Division, Lawrence Berkeley National Laboratory, Berkeley, California 94720, USA }

\date{\today}
\begin{abstract}
We have performed continuous wave and pulsed electron spin resonance measurements of implanted bismuth donors in isotopically enriched silicon-28. Donors are electrically activated via thermal annealing with minimal diffusion. Damage from bismuth ion implantation is repaired during thermal annealing as evidenced by narrow spin resonance linewidths ($B_{pp}\,$=$\,$12$\,\micro$T) and long spin coherence times ($T_{2}\,$=$\,$0.7$\,$ms, at temperature $T\,$=$\,$8$\,$K). The results qualify ion implanted bismuth as a promising candidate for spin qubit integration in silicon. 
\end{abstract}

\maketitle


Electron and nuclear spins of donor atoms in silicon are excellent qubit candidates for quantum information processing \cite{Kane:1998p636,LaddQuantumComputerReview}. Isotope engineered substrates provide a nuclear spin free host environment, resulting in long electron and nuclear spin coherence times of several seconds \cite{MortonElectronNuclearSpinTransfer2008,TyryshkinNatMatOnline}. Spin properties of donor qubit candidates in silicon have been studied mostly for phosphorous and antimony \cite{MortonElectronNuclearSpinTransfer2008,TyryshkinNatMatOnline,Schenkel100PercentDonorAct0.75msAPL2006,MorelloNatureSingleShotElectronSpinReadout2010}. Bismuth donors in silicon are unique in exhibiting a relatively large zero field splitting of 7.4$\,$GHz. Thus, they have attracted attention as potential nuclear spin memory and spin qubit candidates \cite{GeorgeBiPRL2010, MorleyBiNatMat2010} that could be coupled to superconducting resonators \cite{GeorgeBiPRL2010,HatridgeSQUIDPRB2011,VijayNanobridgeAPL2010}. Bismuth is the deepest donor in silicon with a binding energy of 70$\,$meV and a corresponding small Bohr radius. The small Bohr radius and bismuth's reduced effective gyromagnetic ratio \cite{GeorgeBiPRL2010} can make it less susceptible to interface noise at a given implant depth and make bismuth very desirable for quantum logic implementation via magnetic dipolar coupling \cite{ deSousaMagneticDipolarCouplingPRA2004}. Furthermore, bismuth is also the heaviest donor in silicon and thus shows the least ion range straggling during ion implantation, which enables for donor qubit placement with high spatial resolution \cite{SRIM2008,SchenkelDonorDotArxiv2011}.

To date, studies of spin resonance properties of bismuth in silicon have been performed with bulk doped natural silicon \cite{GeorgeBiPRL2010, MorleyBiNatMat2010,Belli-pESR-PRB2011,MohammadyBiQubitsSiPRL2010,SekiguchiThewaltBiPRL2010} whereas silicon-28 material is preferable for improved spin coherence properties. Electrical activation of implanted bismuth via thermal anneals has been studied for relatively high implant doses \cite{YamamotoBiHall2001,AbramofBiTransportPRB1997,daSilvaBiActivationJAP1996,deSouzaBiActAnnealJAP1993,TavakoliBiSPER-APL2005}, and concentrations close to the the metal-insulator transition ($N_c\,$=$\,1.7\,\times\,$10$^{19}\,$cm$^{-3}$) \cite{AbramofBiTransportPRB1997}. High implant doses ($\apprge\,$1$\,\times\,10^{14}\,$cm$^{-2}$ for keV Bi-ions at room temperature) amorphize the silicon lattice \cite{MayerIonImplantation}. Solid-phase epitaxial regrowth (SPER) can be used to prevent diffusion of bismuth atoms and incorporate them on substitutional sites. This can lead to electrically active concentrations well above the low relatively solubility limit of bismuth in silicon (which is, e.g. 2.3$\,\times\,$10$^{17}\,$cm$^{-3}$ at 1150$\,\celsius$ \cite{TrumboreSolubilityLimitsBellSystTechJ1960}). For SPER, thermal anneals at low temperatures, e.g. few minutes at 600$\,\celsius$ yield electrical activation levels of up to 90$\,\%$ \cite{deSouzaBiActAnnealJAP1993}. For low implant doses and dopant concentrations, which are desirable for long spin coherence times, no amorphization of the silicon crystal occurs during ion implantation. Thus, the electrical activation levels and annealing conditions will deviate from those achieved with SPER. Compared to other donors, the high atomic mass of bismuth (atomic weight$\,$=$\,$209) results in an increased defect density during the implantation process. This raises the question whether this intense implantation damage can be repaired effectively during activation anneals to achieve long spin coherence times and high electrical activation yields (EAY) with minimal diffusion. Here, we report on the formation of bismuth-doped silicon-28 by ion implantation, electrical activation of implanted donors and characterization of their spin coherence properties via continuous (cw) and pulsed electron spin resonance (ESR) measurements.

Float zone silicon wafers with a resistivity \textgreater$\,$10,000$\,\Omega$cm and a natural isotope abundance are used for the donor activation study via Hall measurements. Bismuth-209 is implanted at room-temperature under tilt angle of $\,7\degree$. A total fluence of 1.1$\,\times\,$10$^{12}\,$cm$^{-2}$ is implanted at kinetic energies of $E_{kin}\,$=$\,40, 80, 120, 200$ and 360$\,$keV resulting in a peak concentration of 9$\,\times\,$10$^{16}\,$cm$^{-3}$ between a depth of 20 and 150$\,$nm. A box like implant profile was chosen in order to maximize the number of implanted bismuth atoms while keeping the peak concentration below 1$\,\times\,10^{17}\,$cm$^{-3}$. Implant profiles before and after thermal annealing are studied via secondary ion mass spectroscopy (SIMS) measurements \cite{EAG}. The samples are annealed with an AG Associates Heatpulse 210 rapid thermal annealer in nitrogen atmosphere. Unimplanted samples from the same material are annealed at each annealing temperature as controls for sheet carrier density measurements carried out with a Hall effect measurement system (ECOPIA HMS-3000) at room temperature in a dark ambient. EAY values are calculated from the difference of carrier densities in implanted versus unimplanted samples and normalized by the total implant dose. Samples for electron spin coherence measurements consist of a $700\:$nm isotopically enriched $99.95\,\%\:$silicon-28 epitaxial (epi) layer on natural silicon (100) substrate.
\begin{figure}[t]
\includegraphics[width=3.3in]{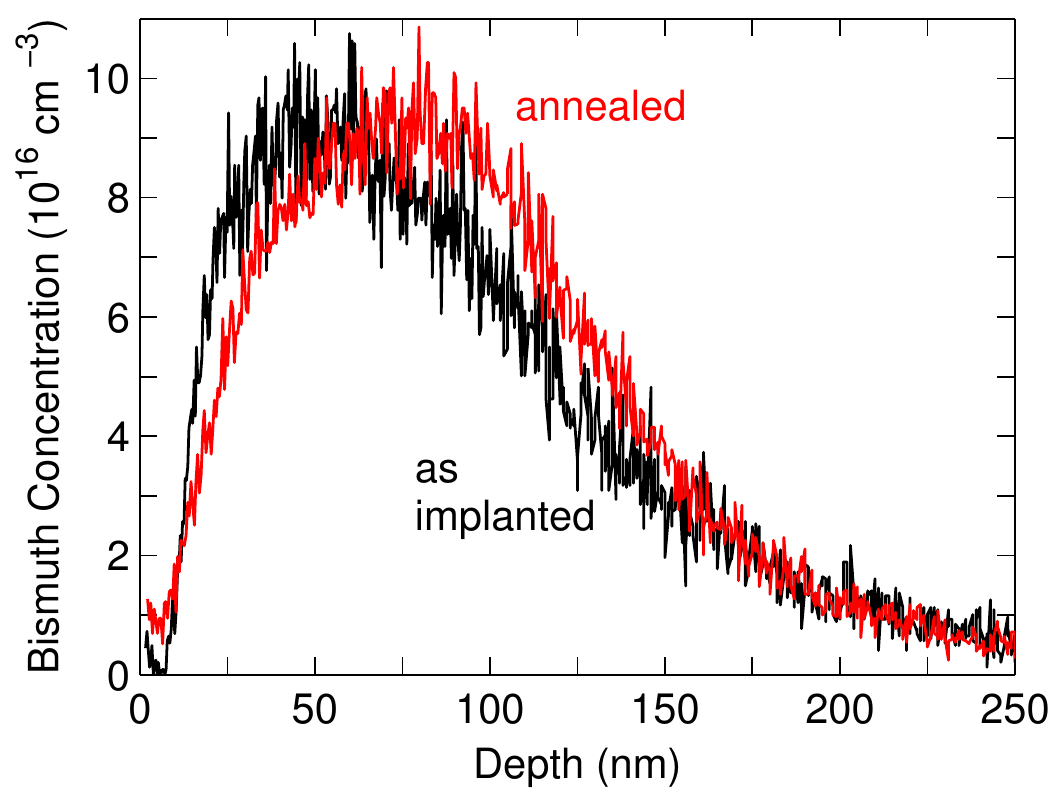}
\caption{\label{fig_SIMS}{SIMS measurements of implanted bismuth-209 in isotopically enriched silicon-28 with a native oxide layer. The total implant dose is 1.1$\,\times\,$10$^{12}\,$cm$^{-2}$ with implant energies ranging from $E_{kin}\,$=$\,$40, 80, 120, 200 to 360$\,$keV. The profiles show the dopant concentrations before (black) and after a thermal activation anneal (red - 20$\,$min at 800$\,\celsius$).}}
\end{figure}
\begin{figure}[b]
\includegraphics[width=3.3in]{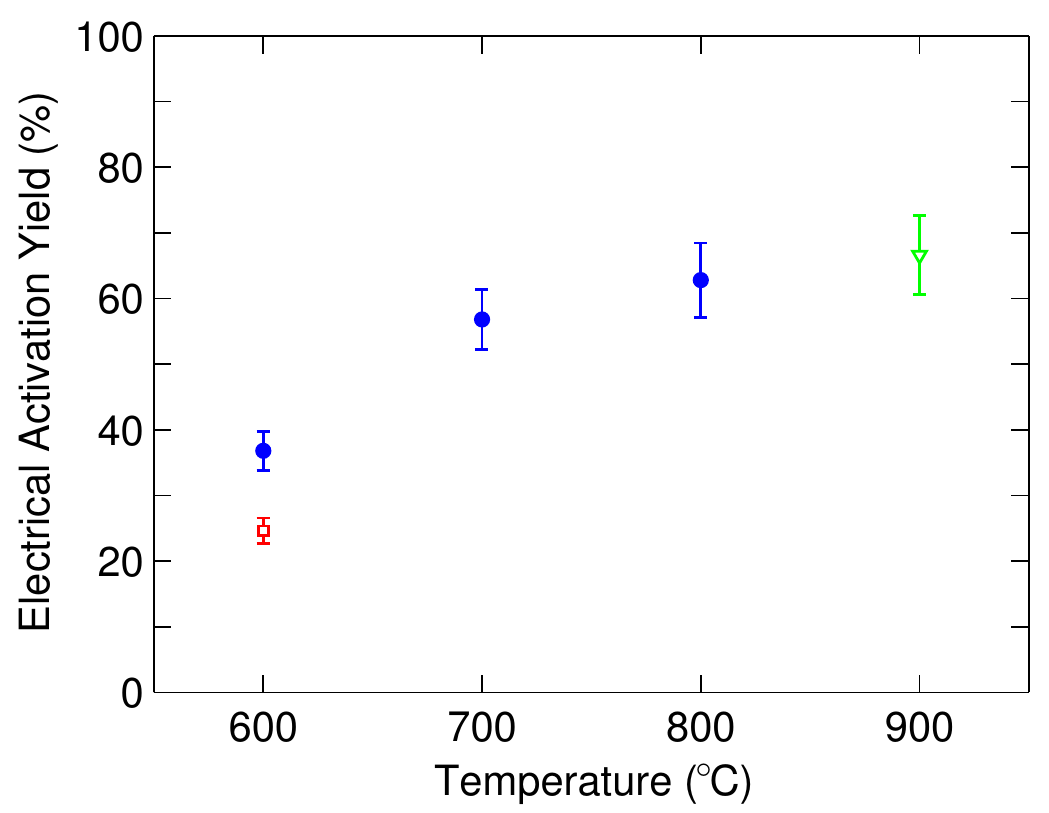}
\caption{\label{fig_HallEAY}{Electrical activation yields of implanted bismuth ion in silicon for a series of annealing conditions; circles (blue): 20$\,$min; square (red): 5$\,$min; triangle (green): 15$\,$min.}}
\end{figure}
Identical bismuth implantation parameters are used as for the activation study and the donors are activated by thermal annealing at 800$\,\celsius$ for 20$\,$min in nitrogen atmosphere. To increase the signal, five samples (total area $\approx\,$1.4$\,$cm$^2$) were stacked in the ESR cavity resulting in 1.5$\,\times\,$10$^{12}$ probed bismuth atoms. ESR measurements are carried out with a Bruker ESP300E X-band EPR spectrometer operating at $f_{\mu w}\,$=$\,$9.42$\,$GHz with a rectangular TE$_{102}$ microwave cavity (for cw-ESR only) and a Bruker ElexSys E680 X-band ($f_{\mu w}\,$=$\,$9.7$\,$GHz) spectrometer with a low temperature helium-flow cryostat (Oxford CF935).

The bismuth concentration profiles from as-implanted and annealed (20$\,$min at 800$\,\celsius$) silicon-28 samples can be seen in Fig.~\ref{fig_SIMS}. The integrated areas from the SIMS measurements are 16$\,\%$ below and 18$\,\%$ above the targeted implantation fluence in the as-implanted and annealed case, respectively. To compensate for this SIMS calibration error the dopant concentrations of both curves are scaled so that the integrated areas match each other and the targeted fluence of 1.1$\,\times\,$10$^{12}\,$cm$^{-2}$. The annealed profile appears to have moved to slightly larger depths. But since the peak value for the annealed profile stays the same and the difference in the full width half maximum is less than 10$\,$nm, we attribute this shift to a depth calibration uncertainty in the SIMS measurements. No bismuth segregation towards the interface is observed as had been observed for higher annealing temperatures \cite{SchenkelNIMB2009}.

Electrical activation levels increase with annealing temperature and reach a value of 67$\,\%$ for annealing at 900$\,\celsius$ (15 min). This trend is consistent with results reported by \cite{YamamotoBiHall2001} for similar bismuth concentrations but it is in contrast to results for high dose bismuth implants which were activated via the SPER technique \cite{YamamotoBiHall2001,deSouzaBiActAnnealJAP1993,TavakoliBiSPER-APL2005}.
We speculate that further enhancements of EAY will be possible through refinement of annealing recipes, possibly together with defect engineering, e.g. through pre-amorphization implants.

\begin{figure}[t]
\includegraphics[width=3.3in]{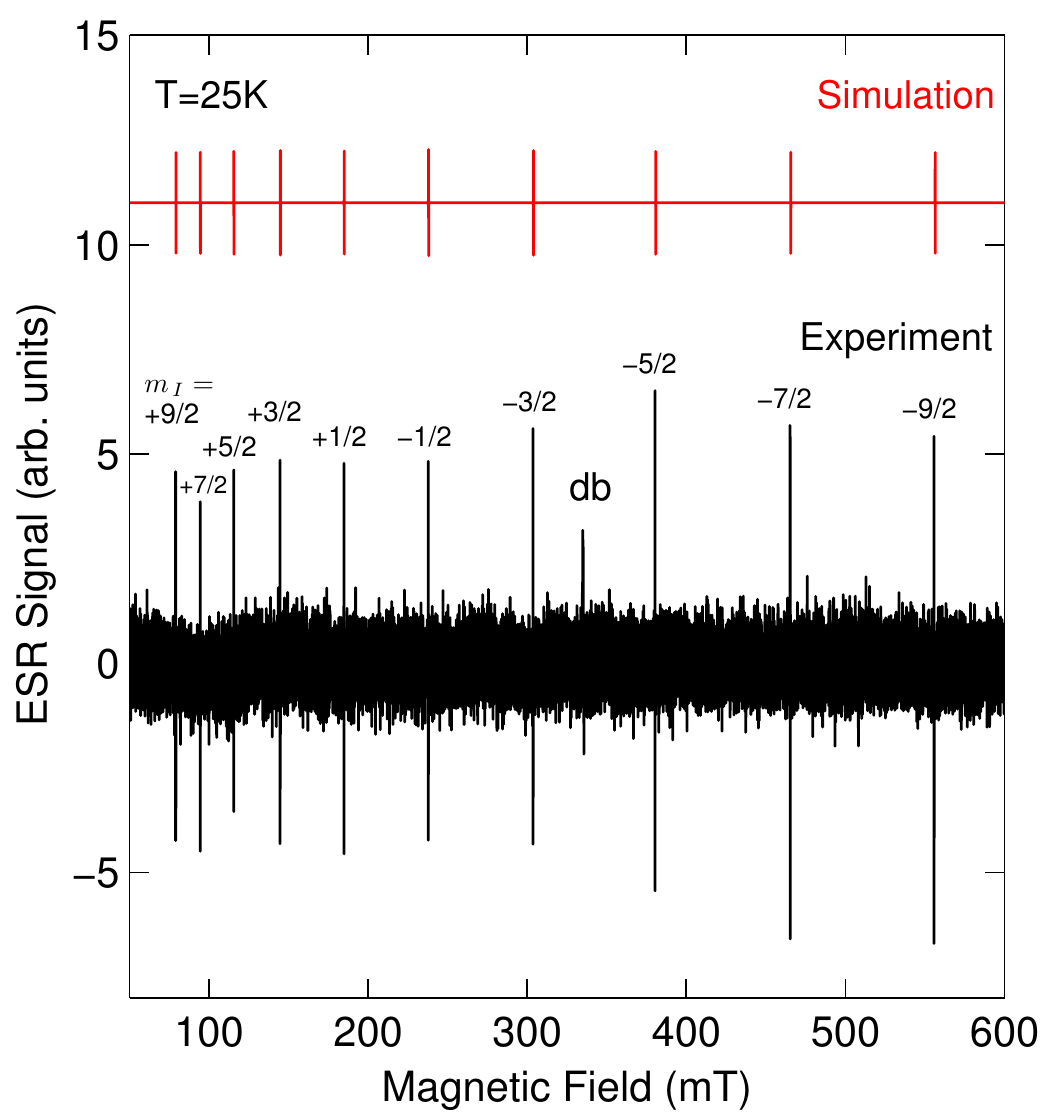}
\caption{\label{fig_cwESR}{Simulated cw-ESR spectrum of Si:Bi (top) showing the expected line positions for a $\mu$-wave frequency of 9.42$\,$GHz and data (bottom) measured at $T=25\:$K. The line at $B\approx\,$335.4$\,$mT originates from dangling bonds at the silicon surface.}}
\end{figure}
\begin{figure}[b]
\includegraphics[width=3.3in]{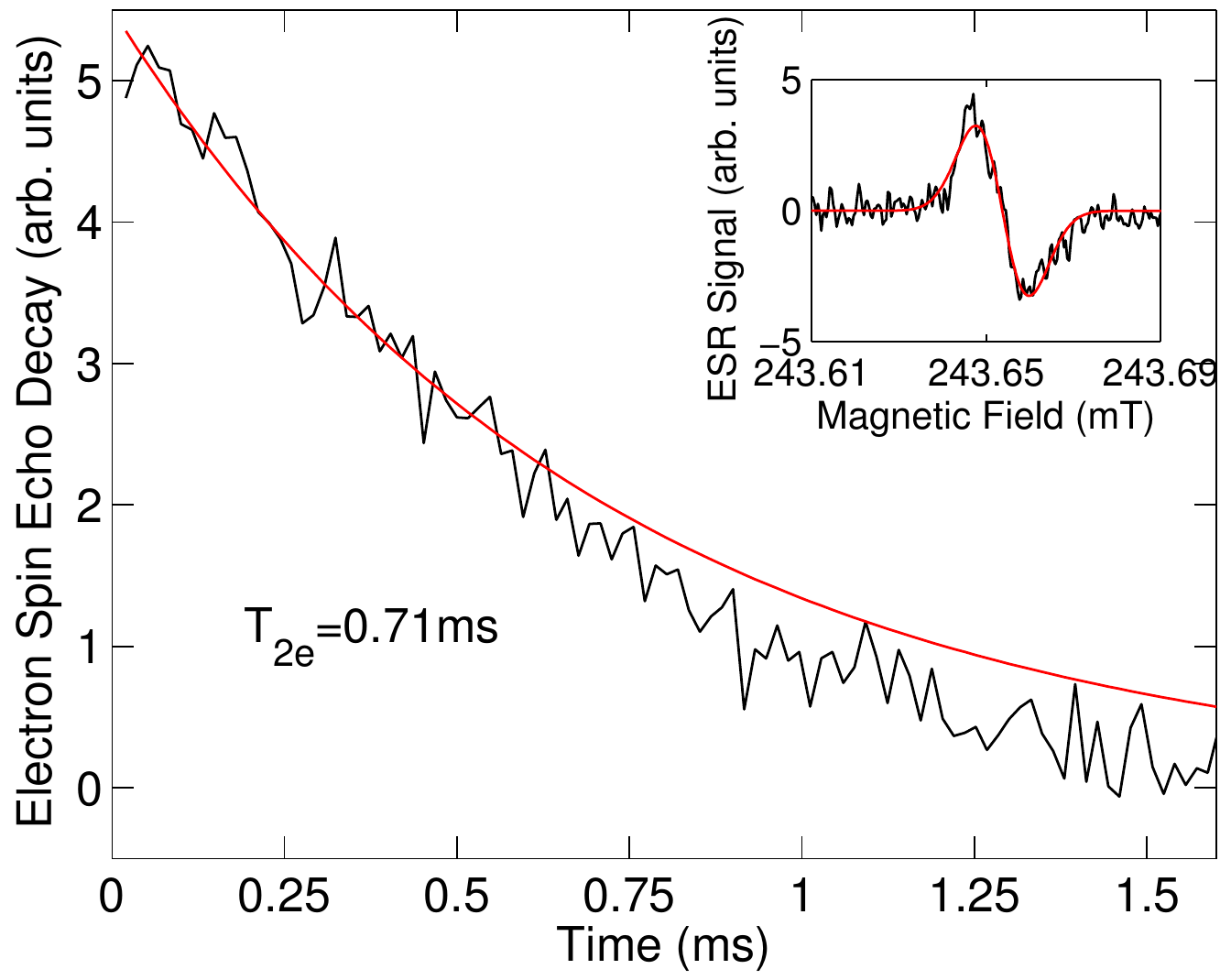}
\caption{\label{fig_T2}{Electron spin echo (ESE) decay using two axis refocusing pulses (XYXY) measured at the $m_I$$\,$=$\,$-1/2 line of $^{28}$Si:Bi at $T\,$=$\,8\:$K. The decay signal at times longer than 0.5$\,$ms is distorted by the phase noise resulting from magnetic field fluctuations \cite{TyryshkinCoherenceOfSpinsSiliconJPCM2006}, and therefore only the initial portion of the decay is used in an exponential fit to extract the electron spin coherence time of $T_{2e}\,$=$\,$0.71$\,\pm\,$0.08$\,$ms. The inset displays a cw-ESR spectrum of the $m_I$$\,$=$\,$-1/2 line and its fit yields a peak-to-peak line-width of 12.2$\,\pm\,$0.4$\,\micro$T.}}
\end{figure}

Due to the large nuclear spin and hyperfine interaction ($I\,$=$\,$9/2, $A\,$=$\,$1475.4$\,$MHz) \cite{FeherESRENDOR1959}, the ESR spectrum stretches across a magnetic field range of 0.5$\,$T at X-band. Fig.~\ref{fig_cwESR} shows simulations of the expected line positions \cite{StollEasySpin2006} and our cw-ESR data at $T\,$=$\,$25$\,$K. All lines match the predicted field positions verifying the successful implantation and activation of bismuth into silicon-28. An additional line is visible at $B\approx\,$335.4$\,$mT which results from dangling bonds (db) at the silicon surface \cite{StesmansPbJAP1998}. A narrow field sweep across the $m_I$$\,$=$\,$-1/2 line at $T\,$=$\,$8$\,$K and $f_{\mu w}\,$=$\,$9.53$\,$GHz is shown as an inset in Fig.~\ref{fig_T2}. The spectrum is taken at a microwave power $P_{\mu w}\,$=$\,$2$\,\micro$W and close to saturation of the peak-to-peak signal amplitude with a modulation amplitude $B_{mod}\,$=$\,$10$\,\micro$T. The Gaussian line fit yields a peak-to-peak line-width $B_{pp}\,$=$\,$12.2$\,\pm\,$0.4$\,\micro$T. Electron spin echo (ESE) decay measurements at $T\,$=$\,$8$\,$K using two axis refocusing pulses (XYXY) are used to determine electron spin coherence times for the $m_I$$\,$=$\,$-1/2 hyperfine line. The fit of the entire ESE signal to a simple exponential decay yields a spin coherence life time of $T_{2e}\,$=$\,$0.57$\,\pm\,$0.03$\,$ms. However, this number can be viewed only as a lower bound of the coherence life time because the decay curve is distorted by the phase noise at times longer than 0.5$\,$ms \cite{TyryshkinCoherenceOfSpinsSiliconJPCM2006}. A more accurate estimate of $T_{2e}\,$=$\,$0.71$\,\pm\,$0.08$\,$ms can be obtained by fitting only the initial potion of the decay (least distorted by the phase noise) as shown with the red curve in Fig.~\ref{fig_T2}. The observed electron spin coherence times are consistent with mechanisms of instantaneous diffusion due to nearby bismuth donors \cite{TyryshkinNatMatOnline} at the peak concentration of 9$\,\times\,10^{16}\,$cm$^{-3}$ (or 9$\,\times\,10^{15}$ spins per resonance line /$\,$cm$^3$). The value of $T_{2e}\,$=$\,$0.7$\,$ms is similar to T$_{2e}$ found for implanted Sb donors at a lower concentration of 3$\,\times\,10^{16}\,$cm$^{-3}$ (5$\,\times\,10^{15}$ spins per line /$\,$cm$^3$) in the presence of a hydrogen passivated silicon surface \cite{Schenkel100PercentDonorAct0.75msAPL2006}. The smaller Bohr radius of Bi-donors and its reduced effective gyromagnetic ratio can contribute to a smaller susceptibility to both surface noise at a given implant depth and to decoherence through coupling to neighboring donors at a given concentration \cite{GeorgeBiPRL2010}. This favors bismuth for implementation of quantum logic through magnetic dipolar coupling \cite{deSousaMagneticDipolarCouplingPRA2004}. Spin coherence lifetimes for growth doped phosphorus epi and implanted antimony films in enriched silicon-28 of similar implantation parameters and donor concentrations (per nuclear spin orientation) show electron spin coherence times of 0.3 to 0.75$\,$ms which are comparable to our results.

In conclusion, we report on ion implantation and electrical activation of bismuth-209 in isotopically enriched silicon-28 samples with minimal dopant diffusion. The obtained narrow linewidths and long electron spin coherence lifetimes are comparable to other implanted donor species in silicon. This shows that the intense implantation damage from heavy ion implants  to the host lattice is repaired effectively and does not affect the spin coherence properties negatively. Our results qualify implanted bismuth donors as a very promising candidate for spin qubit integration in silicon.

We thank the UC Berkeley Marvell Nanolab staff for technical support in device fabrication. This work was supported by the US National Security Agency under 100000080295. Additional supports by DOE under contract no DE-AC02-05CH11231 (LBNL), EPSRC through CAESR EP/D048559/1 (Oxford), and NSF through the Princeton MRSEC under Grant No. DMR-0213706 (Princeton) are also acknowledged. V. L. is supported by Konrad-Adenauer-Stiftung e.V., EPSRC DTA and Trinity College Oxford. J.J.L.M. is supported by The Royal Society and St. John's College, Oxford.
%

\end{document}